\documentclass[doublecolumn]{aa}

\usepackage[utf8]{inputenc}
\usepackage{graphicx}
\usepackage[english]{babel}
\usepackage{natbib,twoopt}
\usepackage[breaklinks=true]{hyperref} 
\bibpunct{(}{)}{;}{a}{}{,}             
\usepackage{amsmath}
\usepackage{placeins}
\usepackage{mathtools}
\usepackage{amssymb}
\usepackage{txfonts}
\usepackage{color}
\usepackage{rotating}
\usepackage{enumitem}
\usepackage[labelfont=bf]{caption}
\usepackage{multirow}
\usepackage{booktabs}
\usepackage{subcaption}
\usepackage{ragged2e}
\usepackage{threeparttable}
\usepackage{url}
\hypersetup{colorlinks=true,citecolor=blue}
\usepackage{footnotebackref}

\begin{document}

\title{First-principles calculations of far-wing absorption in He-broadened CO$_2$ infrared spectra}

\author{
  Ha Tran\inst{}
  \and
  Emile Ducreux\inst{}
}

\institute{
Laboratoire de Météorologie Dynamique, IPSL, Sorbonne Université, École Normale Supérieure, Université PSL, École polytechnique, Institut Polytechnique de Paris, CNRS, Paris, France\\
\email{htran@lmd.ipsl.fr}
}

  \date{Submitted in A\&A}
  \abstract{We report theoretical predictions and modeling of far-wing absorption of CO$_2$ broadened by He over a wide temperature range (200--1000 K), relevant to exoplanet atmosphere studies. Accurate modeling of He-broadened CO$_2$ absorption is essential for opacity calculations used to characterize the composition, structure, and climate of exoplanet atmospheres. In particular, the strong CO$_2$ band at 4.3 $\mu$m has been clearly identified in the spectra of several exoplanets observed with the James Webb Space Telescope (JWST). Requantized classical molecular dynamics simulations were first performed to predict absorption in the far wings of the 4.3 $\mu$m band of CO$_2$. The resulting predictions show excellent agreement with available experimental data at room temperature. The theoretical predictions and measured data were then used to parameterize an impact energy-corrected sudden model, which allows us to accurately reproduce the far-wing absorption of He-broadened CO$_2$ in several infrared spectral regions over the (200--1000) K temperature range. Finally, for applications to exoplanet atmosphere opacity calculations, we derived temperature-dependent $\chi$-factor corrections that account for the sub-Lorentzian behavior of the far wing of the major infrared bands of CO$_2$.}


\titlerunning{Far-wing absorption of He-broadened CO$_2$ infrared lines}
\authorrunning{Tran \& Ducreux} 

   \maketitle
\section{Introduction}

Carbon dioxide (CO$_2$) is a key absorber in exoplanet atmospheres, particularly in gas giants, for which the 4.3 $\mu$m CO$_2$ absorption band (i.e., the $\nu_3$ band) was clearly detected in spectra obtained with the James Webb Space Telescope (JWST) (e.g., \citealt{rustamkulov_early_2023,carter2024_JWST,Balmer_2025,fu2025_JWST}). Accurate knowledge of CO$_2$ absorption, from the band center to the far wings, is essential not only for the characterization of exoplanet atmospheres but also for modeling their thermal structure and climate, which depend strongly on the opacity in the deeper atmospheric layers.

One of the main limitations in interpreting observations and in climate modeling lies in the accuracy of atmospheric opacity calculations, mainly due to the limited knowledge of pressure broadening and far-wing absorption together with their temperature dependence \citep{niraula_impending_2022, chaverot2023first}. In contrast to terrestrial atmospheric conditions, relatively few laboratory studies investigated the line shape of CO$_2$ broadened by the dominant constituents of giant exoplanet atmospheres, namely H$_2$ and He \citep{WILZEWSKI_HITRAN_2016,tan_h2_2022}. In addition, the temperatures relevant to exoplanet atmospheres span a very broad range, from about 100 K to 3000 K \citep{fortney2019need}, well beyond the limits accessible to most laboratory experimental setups.

Recent efforts have focused on accurate measurements and calculations of the H$_2$ and He pressure-broadening coefficients of CO$_2$ lines at temperatures relevant to exoplanet atmospheres (see \citealt{HENDAOUI_JQSRT_CO2He_2025,hendaoui2026_CO2H2, HENDAOUI_CMDS_CO2He_2026} and references therein). These data are expected to significantly improve the accuracy of opacity calculations for H$_2$- and He-rich exoplanet atmospheres. However, only a limited amount of information is available for the far line wings, restricted to narrow spectral regions and room temperature. It is now well established that absorption in the far wings is significantly weaker than predicted by a Lorentzian profile \citep{hartmann_collisional_2021}.

 To our knowledge, only a few studies have investigated far-wing absorption of CO$_2$-He mixtures \citep{Burch_JOSA_1969, baranov1981_CO2He_edges, dokuchaev1985_CO2, thibault1994_CO2He, ozanne1995_CO2He}. \citet{Burch_JOSA_1969} measured absorption in the far-wing regions of the $\nu_1+\nu_3$ band between 3770 and 3860 cm$^{-1}$ and of the $3\nu_3$ band between 6990 and 7002 cm$^{-1}$ at room temperature. From these measurements, they introduced a band- and wavenumber-dependent factor, $\chi$, smaller than unity, to describe the observed sub-Lorentzian behavior.  \citet{dokuchaev1985_CO2} measured CO$_2$ absorption in mixtures with He, Ne, Ar, Xe, H$_2$, D$_2$, and N$_2$ at 294 K in the 2390--2465 cm$^{-1}$ spectral region and showed that the far-wing absorption in a CO$_2$--He mixture is significantly weaker than that in a CO$_2$--N$_2$ mixture, in agreement with the findings of \citet{Burch_JOSA_1969}. \citet{ozanne1995_CO2He} measured far-wing absorption in the $\nu_3$ and $3\nu_3$ bands, in the 2400--2500 cm$^{-1}$ and 6990--7010 cm$^{-1}$ spectral regions respectively, at room temperature and high densities. Their results were consistent with those of \citet{dokuchaev1985_CO2} in the $\nu_3$ band wings.  

The sub-Lorentzian behavior observed in the far wings is due to collisional interference between lines (line mixing) and to the breakdown of the impact approximation. Interactions between He atoms and molecules are generally weak, with a steep repulsive wall and a small well that rapidly vanishes at larger intermolecular distances. As a result, collision durations are much shorter than the mean time between successive collisions \citep{tran2023forein_HCl,reed2023effect,Tran_O2_depletion}, suggesting that non-impact effects should be negligible in He-broadened CO$_2$ spectra. \citet{ozanne1995_CO2He} successfully used an impact Energy-Corrected Sudden (ECS) model to describe line-mixing effects in CO$_2$-He absorption spectra at room temperature and high densities. This model enables the calculation of collisional line interferences from rotational state-to-state transfer rates. Good agreement with measurements was obtained from the band centers to the far wings in both the $\nu_3$ and $3\nu_3$ bands. Similar approaches were applied to model line-mixing effects at the band centers and in the near wings for various spectral regions \citep{Boissoles_CO2He_JCP_1994,boissoles_CO2He_JQSRT_1996,khalil_CO2He_CPL_1996,rodrigues_CO2He_JCP_1998,hartmann_CO2_LM_JCP_1999,boulet_CO2He_JCP_1999}. These studies, however, were limited to room temperature only, likely because both measured spectra and line-broadening coefficients required to constrain the models were only available under such conditions. 

More recently, classical molecular dynamics simulations (CMDS) were used to predict far-wing absorption for pure CO$_2$ and CO$_2$--H$_2$O mixtures, with very good agreement with measurements \citep{Hartmann_JCP_2010, Hartmann_CMDS_CO2H2O_2018}. This method, based on the use of accurate intermolecular potentials and Newtonian equations of motion, enables the computation of spectral line profiles from the line center \citep{Nguyen_CO2N2_JCP_2018, NGUYEN_CMDS_CO2_2020, NGO_Tran_CMDS_CO2_2025, TRAN_CMDS_CO2N2_2025} to the far wings without the use of adjustable parameters. CMDS were also successfully applied to predict pressure-broadening parameters over wide temperature ranges for various species, including He- and H$_2$-broadening coefficients of CO$_2$ lines \citep{hendaoui2026_CO2H2, HENDAOUI_CMDS_CO2He_2026}. 

In this work, we use CMDS to predict far-wing absorption of the $\nu_3$ band of He-broadened CO$_2$ over a wide temperature range, from 200 K to 1000 K. The predictions are compared with experimental data available at room temperature. The CMDS results are then used to parameterize the temperature dependence of the impact ECS model. Indeed, the latter is based on the use of some empirical parameters to calculate the rotational state-to-state transfer rates. These empirical parameters are obtained from fits of far-wing absorption predicted by CMDS at various temperatures. Once validated, this ECS model enables calculations of far-wing absorption for He-broadened CO$_2$ in various infrared bands over a broad temperature range. For practical applications, the resulting far-wing absorptions are finally fitted using temperature-dependent $\chi$-factor corrections suitable for atmospheric radiative transfer and opacity calculations. 

The paper is organized as follows. Section~\ref{sec:cmds} presents the CMDS calculations, while Sect.~\ref{sec:ecs} describes the ECS model and the associated data used. Comparisons between modeled absorptions, CMDS predictions, and available measurements are presented in Sect.~\ref{sec:comparisons}. In Sect.~\ref{sec:chi}, we derive the $\chi$-factor parameterization of the far-wing absorption computed with the ECS model and illustrate its impact on opacity calculations. Conclusions are given in Sect.~\ref{sec:conclusions}. 

\section{CMDS}
\label{sec:cmds}
We first computed far-wing absorption spectra of CO$_2$ broadened by He using the requantized CMDS (rCMDS) approach described in  \citet{HENDAOUI_CMDS_CO2He_2026}. That study demonstrated that He-broadening coefficients of CO$_2$ lines predicted by rCMDS are in very good agreement with experimental data over a wide temperature range, from 123 K to 760 K. The simulation methodology, based on the formalism described in chapter 3 of \citet{Allen-Tidesley}, was detailed in our previous studies devoted to the prediction of pressure-broadening coefficients of CO$_2$ lines with various collision partners \citep{NGO_Tran_CMDS_CO2_2025,TRAN_CMDS_CO2N2_2025,hendaoui2026_CO2H2,HENDAOUI_CMDS_CO2He_2026}. In the present work, simulations were performed at four temperatures: 200 K, 296 K, 500 K, and 1000 K.

Mixtures containing 50\% CO$_2$ and 50\% He were considered, but only CO$_2$--He interactions were included in order to reduce the computational cost. Since the contribution of three-body (and higher) interactions is negligible, this approximation is equivalent to simulating CO$_2$ infinitely diluted in He at a total pressure equal to the He partial pressure. The He pressures considered were 10 atm at 200 K and 296 K, 20 atm at 500 K, and 60 atm at 1000 K. For each temperature and pressure condition, we simulated a large number of particles ($80 \times 10^6$) initially placed in cubic boxes with periodic boundary conditions. The position of the center of mass, translational velocity directions, angular momenta, and molecular orientations (for CO$_2$) were randomly initialized, while translational and rotational speeds were sampled from Maxwell--Boltzmann distributions. Using an accurate intermolecular potential \citep{korona2001spectroscopic, HENDAOUI_CMDS_CO2He_2026} and Newton's equations of motion, the time propagation of the system was computed at each time step, up to a maximum simulation time $t_{\max}$. This procedure enables the calculation of the autocorrelation function (ACF) of the dipole moment (along the molecular axis for a parallel band of CO$_2$). The Fourier--Laplace transform of the ACF directly yields the spectral density, or, equivalently, the band-normalized spectral profile.

A requantization procedure was then applied to convert the classical angular momenta into quantum rotational states. To do so, we used the same approach as in \citet{HENDAOUI_CMDS_CO2He_2026}, which was shown to provide very good predictions of He-broadening coefficients over a wide temperature range. In addition, the distinction between the $P$- and $R$-branch transitions was introduced by considering the sign of the angular momentum along the $x$ axis: positive values were assigned to the $R$ branch, whereas negative values were assigned to the $P$ branch. Finally, the requantized angular momentum values used in the ACF calculations were directly linked to the relative positions (with respect to the band center) of the strong $\nu_3$ band lines at 4.3 $\mu$m. The resulting rCMDS spectrum therefore reproduces the structure of the CO$_2$ $\nu_3$ band and can be directly compared with measured absorption profiles.

To compare the rCMDS predictions with the measurements, we scaled the spectra using the integrated $\nu_3$-band intensity obtained from line-by-line calculations based on the HITRAN molecular spectroscopic database \citep{gordon_hitran2020_2021} and, for temperatures above 500~K, the HITEMP high-temperature spectroscopic database \citep{rothman_hitemp_2010}. We note that a direct comparison between the rCMDS-predicted absorption for the $\nu_3$ band and measurements in the 2400--2600 cm$^{-1}$ spectral region is possible because the contributions from weaker bands are negligible there. 
 
\begin{figure}[htbp]
  \centering
  \includegraphics[width=0.5\textwidth]{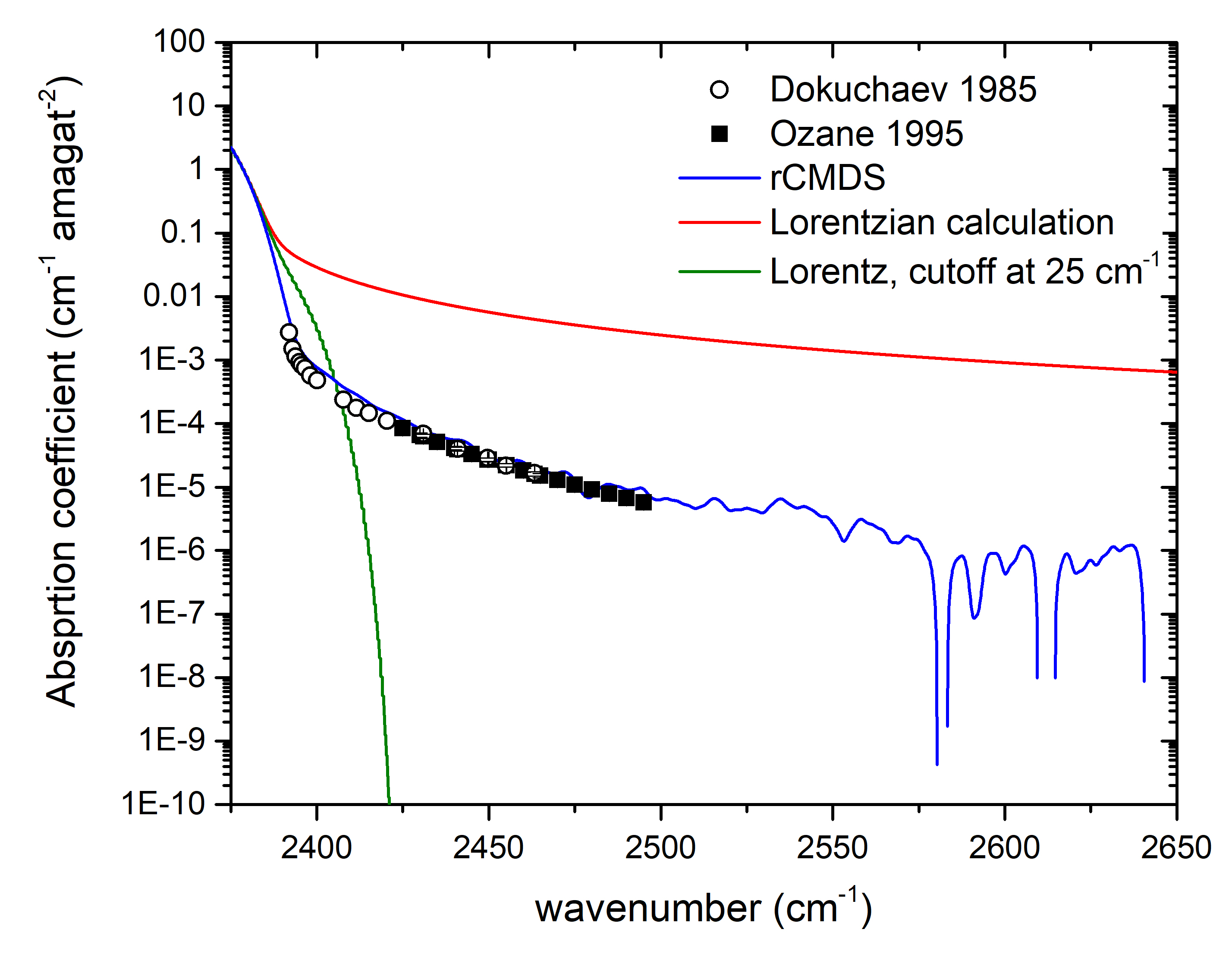}
  \caption{Far-wing absorption coefficients of the CO$_2$ $\nu_3$ band broadened by He, calculated using rCMDS at room temperature and 10 atm of He. The results are compared with  experimental values from \citet{dokuchaev1985_CO2} and \citet{ozanne1995_CO2He}, and with absorption coefficients calculated using the Lorentz model. Results obtained by applying a line-wing cutoff at 25 cm$^{-1}$ are also shown.}
   \label{fig.1}
\end{figure}

Figure~\ref{fig.1} presents the results obtained at 296 K. The rCMDS results exhibit residual oscillations originating from two main effects. First, despite the large number of simulated molecules, the computed ACF is not perfectly smooth and still contains short-time fluctuations associated with individual collision events. These fluctuations become increasingly visible in the far-wing regions, where absorption is very weak. Second, the oscillations observed in the rCMDS results are also due to the limited time extent of the calculations of the ACF. This introduces a convolution of the true absorption spectrum with a cardinal sine function, which also produces oscillatory structures in regions of weak absorption. To reduce these artifacts, an apodization procedure similar to that commonly used in Fourier-transform spectroscopy was applied. Specifically, the ACF was multiplied by a Gaussian window function that remains close to unity while the ACF is significant and smoothly decreases to zero as the maximum simulation time $t_{max}$ approaches. The fluctuations are significantly reduced by apodization, although they are not completely eliminated, particularly in the very far wings (Fig.~\ref{fig.1}), where the absorption is extremely weak.

As shown in Fig.~\ref{fig.1}, the rCMDS predictions are in very good agreement with experimental values of \citet{dokuchaev1985_CO2, ozanne1995_CO2He}. They clearly show large deviations from absorption calculated using simple Lorentzian profiles. It is important to emphasize that no adjustable parameters were introduced in the rCMDS calculations. These results demonstrate that these first-principles calculations are capable of accurately predicting far-wing absorption of CO$_2$ broadened by He. For comparison, Fig.~\ref{fig.1} also shows the results obtained with the Lorentzian line-shape model but imposing a line-wing cutoff at 25 cm$^{-1}$, an assumption commonly adopted in opacity databases and retrieval codes used for exoplanet atmosphere studies, including TauREx, NEMESIS, CHIMERA \citep{barstow2020comparison} and MAESTRO \citep{gharib-nezhad_impact_2024}. As can be seen, this simplified treatment significantly underestimates the far-wing absorption of CO$_2$ broadened by He, highlighting the importance of properly accounting for the non-Lorentzian far-wing behavior in opacity calculations.

Additional rCMDS calculations were subsequently performed at 200 K, 500 K, and 1000 K. For each temperature and pressure condition, the rCMDS-calculated spectrum of the $\nu_3$ band was scaled to have the same integrated band intensity as the corresponding line-by-line spectrum. These spectra were then used to parameterize the impact ECS model, as described in the following section.

\section{Impact ECS line-mixing model}
\label{sec:ecs}
\subsection{General equations}
To account for line-mixing effects in the He-broadened CO$_2$ spectrum within the impact approximation, we adopt the formalism commonly used in the literature (see \citealt{hartmann_collisional_2021} and references therein).  The absorption coefficient at wavenumber $\sigma$ is expressed as:  

\begin{multline}
\alpha(\sigma) = N \frac{8 \pi^2 \sigma}{3 h c}
\left[1 - e^{-\frac{h c \sigma}{k_B T}}\right]\\
\sum_{l} \sum_{k}
\rho_l(T)\, d_l d_k \,
\operatorname{Im}
\left\{
\left\langle\!\left\langle
k \middle\| \left[\Sigma - L_0 - i N W(T)\right]^{-1}
\middle\| l
\right\rangle\!\right\rangle
\right\}
\label{Eq.1}
\end{multline}

The summations in Eq.~\ref{Eq.1} extend over all transitions $l$ and $k$ of the absorbing gas. Here, $N$ denotes the number density of the absorber, $\rho_l$ is the Boltzmann population fraction of the lower state of transition $l$, and $d$ represents the dipole matrix element. The notation $\left\langle\!\left\langle
k \middle\| \cdot \middle\| l
\right\rangle\!\right\rangle$ indicates a matrix element. $\left\langle\!\left\langle k \middle\| \Sigma \middle\| l \right\rangle\!\right\rangle = \sigma \, \delta_{lk}$ corresponds to the wavenumber at which the absorption coefficient is evaluated; $\left\langle\!\left\langle k \middle\| L_0 \middle\| l \right\rangle\!\right\rangle = \sigma_l \, \delta_{lk}$ corresponds to the position of transition $l$. $W(T)$ is the collisional relaxation matrix. It should be noted that Eq. \ref{Eq.1} neglects Doppler broadening as well as speed dependence and velocity-change effects \citep{ngo_isolated_2013,hartmann_collisional_2021}. 

The matrix $W(T)$ contains all collisional information for the absorbing molecule. The real and imaginary parts of its diagonal elements correspond, respectively, to the pressure broadening, $\Gamma$, and shift, $\Delta$, for each transition \citep{hartmann_collisional_2021}. The off-diagonal elements quantify the collisional coupling between individual lines. More specifically, $\left\langle\!\left\langle k \middle\| W(T) \middle\| l \right\rangle\!\right\rangle$ corresponds to the collisional transfer rate from the states associated with line $l$ to those  associated with line $k$.

The relaxation matrix $W(T)$ satisfies two fundamental relations: the detailed-balance principle and the sum rule. The detailed-balance principle relates pairs of transposed matrix elements according to:

\begin{equation}   
\left\langle\!\left\langle k \middle\| W(T) \middle\| l \right\rangle\!\right\rangle \, \rho_l(T)
=
\left\langle\!\left\langle l \middle\| W(T) \middle\| k \right\rangle\!\right\rangle \, \rho_k(T)
\label{Eq. 2}
\end{equation}
while the sum rule is expressed as: 
\begin{equation}
\sum_{k} d_k^0 \, \left\langle\!\left\langle k \middle\| W(T) \middle\| l \right\rangle\!\right\rangle = 0
\label{Eq.3}
\end{equation}
where $d_k^0$ denotes the rigid-rotor dipole matrix elements \citep{hartmann_CO2_LM_JCP_1999, niro_spectra_2004}. These should not be confused with the quantities $d_k$ in Eq.~\ref{Eq.1}, which are determined from the line intensities.

Equation \ref{Eq.3} provides a direct connection between the off-diagonal and diagonal elements of the relaxation matrix and can therefore be used to constrain the off-diagonal elements from the pressure-broadening coefficients.

\subsection{ECS model}
Several studies modeled the CO$_2$--He relaxation matrix using the ECS scaling law (e.g., \citealt{Boissoles_CO2He_JCP_1994,ozanne1995_CO2He,boulet_CO2He_JCP_1999, boissoles_CO2He_JQSRT_1996,rodrigues_CO2He_JCP_1998}) where detailed descriptions of the model can be found. The ECS formalism enables the calculation of collisional transfer rates while properly accounting for angular momentum transfer during collisions \citep{hartmann_collisional_2021}. In the present work, we follow the specific ECS formulation previously used to model self- and air-broadened CO$_2$ relaxation matrices \citep{niro_spectra_2004,tran_measurements_2011,cole2023temperature}. The real part of $W(T)$ is expressed as: 
    
\begin{equation}
\begin{aligned}
\operatorname{Re} \left\langle\!\left\langle k \middle\| W(T) \middle\| l \right\rangle\!\right\rangle
&= (-1)^{l_{2i} + l_{2f}} (2 j_i' + 1) \sqrt{(2 j_f' + 1)(2 j_f + 1)} \\
&\quad \times \sum_{\substack{L \text{ even} \\ L \neq 0}} 
\begin{pmatrix} j_i & L & j_i' \\ l_{2i} & 0 & -l_{2i} \end{pmatrix} \\
&\quad \times \begin{pmatrix} j_f & L & j_f' \\ -l_{2f} & 0 & l_{2f} \end{pmatrix}
\begin{Bmatrix} j_i & j_f & 1 \\ j_f' & j_f & L \end{Bmatrix} \\
&\quad \times (2L + 1)\, \frac{\Omega(j_i, T)}{\Omega(L, T)}\, Q(L, T)
\end{aligned}
\label{Eq.4}
\end{equation}
where $j_i$ and $j_f$ are the initial and final rotational quantum numbers associated with transition $l$, while $j_i'$ and $j_f'$ correspond to transition $k$. $l_2$ is the vibrational angular momentum quantum number. The symbols $(\ddots)$ and $\{\ddots\}$ denote the Wigner 3J and 6J coefficients, respectively \citep{edmonds1996angular}. The factor $\Omega$ is an adiabaticity correction that accounts for molecular rotation during collisions \citep{depristo1979quantum, hartmann_collisional_2021}

\begin{equation}
    \Omega(j,T)
=
\left(
1 + \frac{1}{24}
\left(
\frac{\omega_{j,j-2}\, d_c}{\bar{v}(T)}
\right)^2
\right)^{-2}
\label{Eq.5}
\end{equation}
where $d_c$ is a scaling length and $\bar{v}(T)$ is the mean relative speed. The quantity $\omega_{j,j-2}$ denotes the frequency spacing between the rotational levels $j$ and $j-2$.

The rotational cross sections $Q(L,T)$ in Eq.~\ref{Eq.4} are given by:

\begin{equation}
   Q(L,T)=
A(T)\,
\bigl[L(L+1)\bigr]^{-\lambda(T)}
\exp\!\left(-\beta \frac{hcE_L}{k_B T}\right)
\label{eq.6}
\end{equation}
where $E_L$ is the rotational energy of level $L$. \\

The model parameters, i.e., $d_c$, $A(T)$, $\lambda(T)$, and $\beta$, in Eqs.~\ref{Eq.5} and \ref{eq.6}, are generally determined from experimental data such as broadening coefficients (using the sum rule (Eq.~\ref{Eq.3})) and measured absorption spectra \citep{Boissoles_CO2He_JCP_1994,rodrigues_CO2He_JCP_1998,niro_spectra_2004,tran_measurements_2011,cole2023temperature}. At room temperature, we evaluated the three sets of ECS parameters reported by \citet{Boissoles_CO2He_JCP_1994} and \citet{rodrigues_CO2He_JCP_1998}, which were obtained from simultaneous fits to pressure-broadening coefficients and He-broadened CO$_2$ absorption spectra. To this end, broadening coefficients calculated using the sum rule (Eq.~\ref{Eq.3}) using each parameter set were compared to the updated and accurate broadening coefficients reported by  \citet{HENDAOUI_JQSRT_CO2He_2025,HENDAOUI_CMDS_CO2He_2026}.

As expected, these different parameter sets were found to yield reasonable fits of the broadening coefficients with comparable accuracy. Consequently, the final parameter set was selected based on its ability to simultaneously provide a satisfactory representation of the broadening coefficients and the best agreement between calculated and measured as well as rCMDS-predicted absorption spectra in the band wing. The retained parameter set corresponds to that of \citet{Boissoles_CO2He_JCP_1994}, with an updated value of the magnitude parameter $A$ in Eq.~\ref{eq.6}. The resulting parameters, listed in Table~\ref{Table1}, provide the best agreement with the experimental data in the 2390--2495 cm$^{-1}$ spectral region, corresponding to the far wing of the $\nu_3$ band (see Fig.~\ref{fig.2}).

To determine the temperature dependence of the ECS parameters, i.e., $A(T)$ and $\lambda(T)$, we used the pressure-broadening coefficients reported by  \citet{HENDAOUI_CMDS_CO2He_2026} together with absorption spectra in the $\nu_3$ band wing computed from the rCMDS at 200, 500, and 1000 K (see Sect.~\ref{sec:cmds}). The temperature dependence of $A(T)$ and $\lambda(T)$ was represented using the power-law model, which is widely used for the line broadening coefficient and adopted in previous ECS studies \citep{rodrigues_CO2He_JCP_1998,cole2023temperature} i.e.: 

\begin{equation}
A(T) = A(296\,\mathrm K)\left(\frac{296}{T}\right)^{n_A}   
\label{eq.7}
\end{equation}
where $A(296\,\mathrm K)$ is determined at 296 K, and $n_A$ scales the parameter to different temperatures $T$. The values of $n_A$ and $n_{\lambda}$ are also provided in Table~\ref{Table1}.   

\begin{table}[h]
	\centering
    \caption{ECS parameters used.}
	\begin{tabular}{ c c c c c c}
        \hline  \hline 
        $d_c$ ($\mathrm{\AA}$)  &  $A(296\,\mathrm K)$ (cm$^{-1}$\,atm$^{-1}$)  &  $n_A$  &   $\lambda(296\,\mathrm K)$  & $n_\lambda$ &  $\beta$\\
         
        \hline 
        0.5 & 0.058  &  0.601  &  1.20  &  0.09  &  0.0   \\
        \hline 
	\end{tabular}

 \label{Table1}
\end{table}

\subsection{Calculation procedure}

Once the model parameters were determined, the real parts of the off-diagonal elements of the relaxation matrix were calculated using Eqs.~\ref{Eq.4}--\ref{eq.7} together with the detailed balance (Eq.~\ref{Eq. 2}). The imaginary parts of the off-diagonal elements were neglected because they do not significantly affect the shape of the calculated spectra \citep{hartmann_collisional_2021}. For the diagonal elements of $W(T)$, we used the He-broadening coefficients and their temperature dependence provided by \citet{HENDAOUI_CMDS_CO2He_2026}. These pressure-broadening coefficients were computed for temperatures up to 3000 K. For the imaginary contribution of the diagonal elements, we used the measured pressure shifts in the $\nu_3$ band from \citet{HENDAOUI_JQSRT_CO2He_2025}, assuming that there is no temperature dependence. 

The relaxation matrix was constructed separately for each vibrational band. We neglected line mixing between transitions belonging to different vibrational levels because collisional population transfer between these levels is expected to be extremely weak \citep{hartmann_CO2_LM_JCP_1999, hartmann_collisional_2021}. It should also be noted that we employed the renormalization procedure described in \citet{niro_spectra_2004}, to ensure that the relaxation matrix exactly satisfies the detailed balance and the sum rule (Eqs.~\ref{Eq. 2} and \ref{Eq.3}). Consequently, the remaining discrepancy between the broadening coefficients derived from the off-diagonal elements of the relaxation matrix through the sum rule (Eq.~\ref{Eq.3}) and those represented by its diagonal elements has a negligible effect on the computed spectra.

We then calculated spectra from Eq.~\ref{Eq.1} using the relaxation matrix constructed with the ECS model. The calculation employed the relaxation matrices described above together with spectroscopic parameters (line positions, intensities, and lower-state energies) taken from HITRAN \citep{gordon_hitran2020_2021} and, for temperatures above 500 K, HITEMP \citep{rothman_hitemp_2010}.

\section{Results and discussion}
\label{sec:comparisons}

\begin{figure}[htbp]
  \centering
  \includegraphics[width=0.5\textwidth]{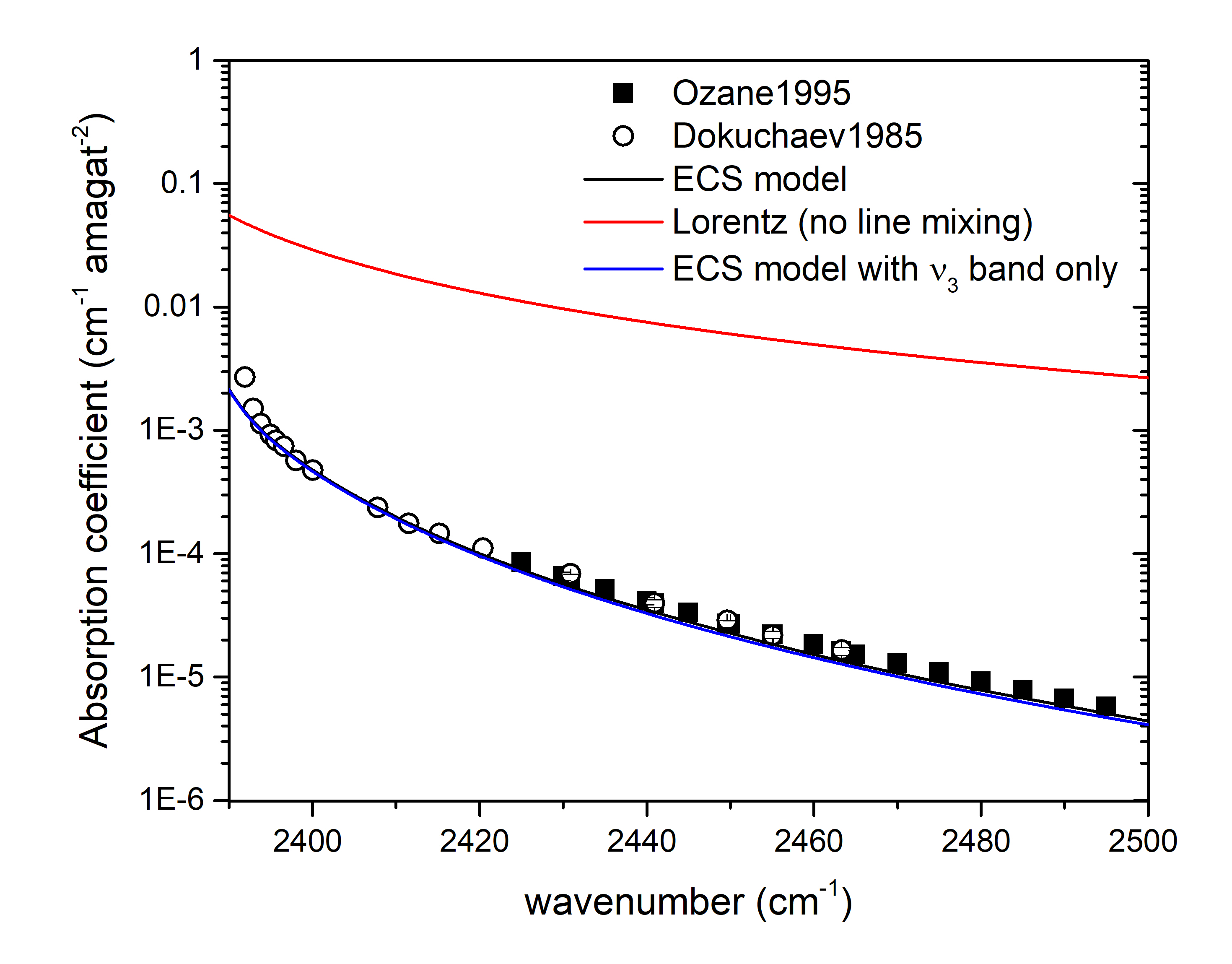}
  \caption{He-broadened CO$_2$ absorption coefficients measured at room temperature in the wing of the $\nu_3$ band \citep{dokuchaev1985_CO2, ozanne1995_CO2He}, compared with calculations performed using the ECS line-mixing model and Lorentzian profiles (without line mixing).}
  \label{fig.2}
\end{figure}

Figure~\ref{fig.2} compares room-temperature absorption coefficients in the 2400--2500 cm$^{-1}$ spectral region calculated with line-mixing effects accounted for using the present ECS parameterization with the experimental data of \citet{dokuchaev1985_CO2, ozanne1995_CO2He}. Contributions from all bands present in the 2000--2800 cm$^{-1}$ spectral range were included in the calculations. The corresponding absorption computed using the simple Lorentzian model, which neglects line-mixing effects, is also shown. As can be seen, the ECS model reproduces the experimental measurements very well over the entire spectral range considered. This result is similar to those reported in Fig.~2 of \citet{ozanne1995_CO2He}, which, however, used a different model formulation. As expected, Lorentzian profiles significantly overestimate far-wing absorption, with discrepancies exceeding two orders of magnitude.

Figure~\ref{fig.2} also presents absorption calculated using only transitions belonging to the strong $\nu_3$ band of the main CO$_2$ isotopologue. The result confirms that, in this spectral region, absorption is dominated by the far-wing contribution of the $\nu_3$ band, while contributions from weaker bands are negligible. Therefore, in this spectral region, rCMDS-computed spectra of the $\nu_3$ band can be directly compared with experimental values. In contrast, in the low-frequency wing of the $\nu_3$ band, for instance, the relative contributions of weaker bands are significantly stronger. Hence, to compare with rCMDS-predicted absorption over the entire spectral region covering the band, ECS calculations were performed considering only the contribution of the $\nu_3$ band.

\begin{figure*}[h]
  \centering
  \includegraphics[width=\textwidth]{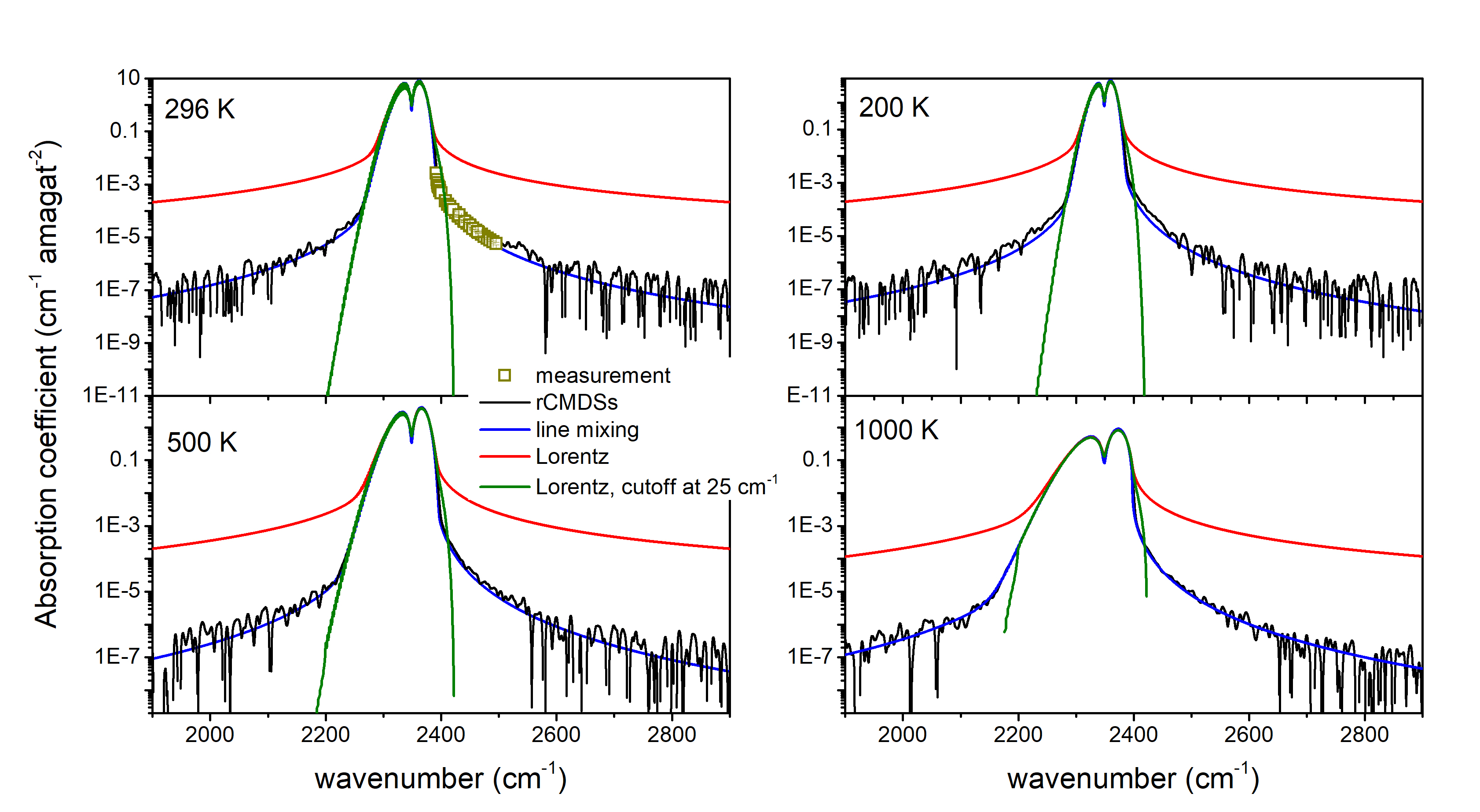}
\caption{Comparison of He-broadened CO$_2$ absorption in the $\nu_3$ band region around 4.3~$\mu$m calculated using rCMDS and the ECS model at 296, 200, 500, and 1000~K. Calculations using purely Lorentzian profiles and Lorentzian profiles with a line-wing cutoff at 25~cm$^{-1}$, together with available experimental measurements \citep{dokuchaev1985_CO2, ozanne1995_CO2He}, are also shown.}
  \label{fig.3}
\end{figure*}

The results obtained at the four temperatures considered in the rCMDS are presented in Fig.~\ref{fig.3}. Excellent agreement is observed between the rCMDS predictions and the spectra computed using the ECS model over both the low- and high-frequency wings of the band, from the near-wing to the far-wing regions. In contrast, the Lorentzian calculations substantially overestimate both the rCMDS and ECS results. The discrepancy increases with increasing distance from the band center, reaching more than three orders of magnitude in the far wings. The 25~cm$^{-1}$ line-wing cutoff calculations, on the other hand, yield far-wing absorption that rapidly decreases to zero and therefore significantly underestimate the rCMDS and ECS results.

It is important to note that rCMDS calculations account for both collisional line-mixing and collision-duration (non-impact) effects. The excellent agreement between the rCMDS predictions and the impact-based ECS model demonstrates that non-impact effects have a negligible influence on the far-wing absorption of CO$_2$ broadened by He. This finding is consistent with previous studies \citep{tran2023forein_HCl, reed2023effect, Tran_O2_depletion}, which showed that non-impact effects of molecule--He collisions are negligible. For systems involving stronger intermolecular interactions, such as CO$_2$--CO$_2$, however, non-impact effects must be taken into account to accurately describe far-wing absorption \citep{tran_measurements_2011}.

\begin{figure}[htbp]
  \centering
  \includegraphics[width=0.5\textwidth]{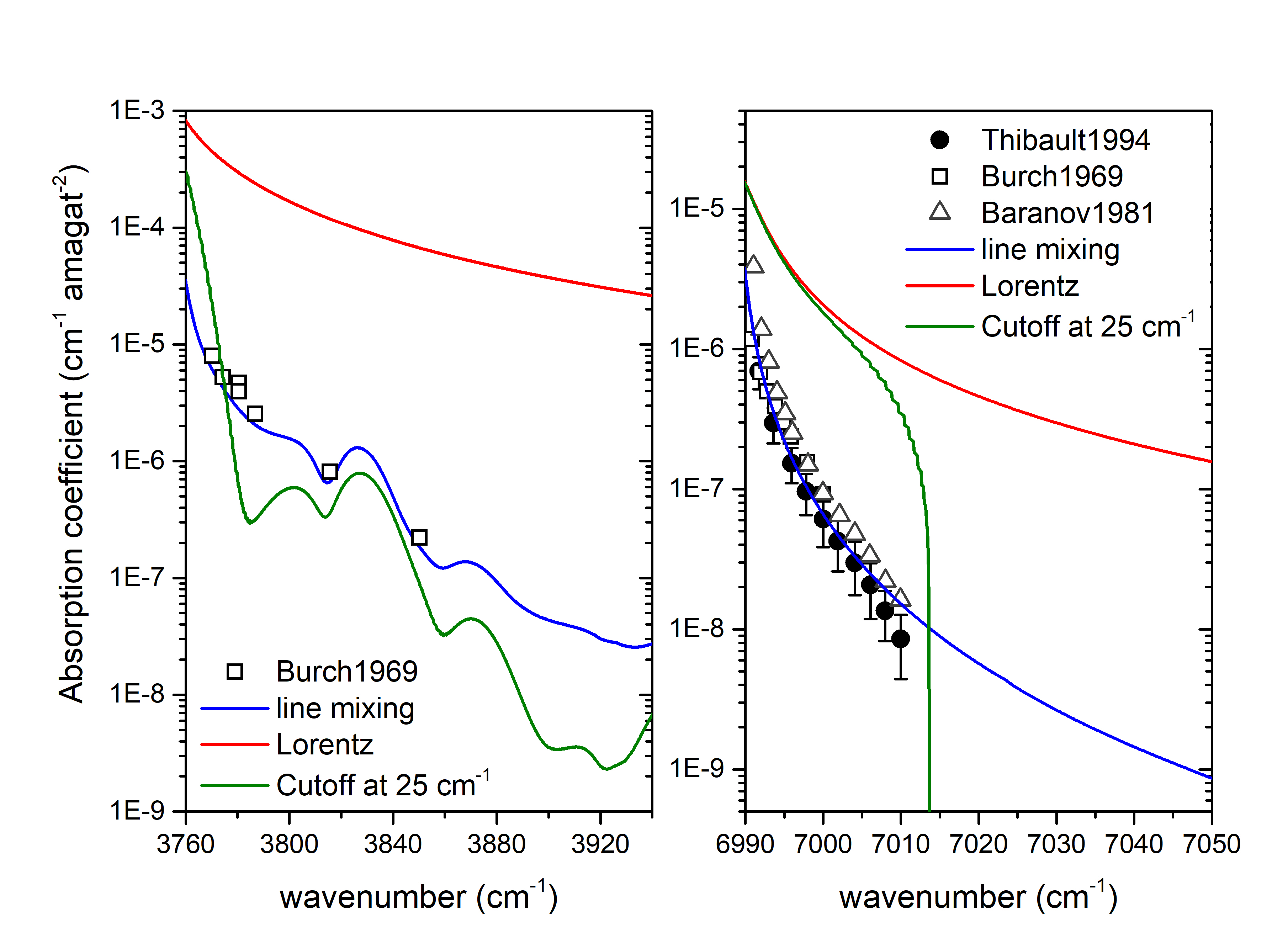}
  \caption{Room-temperature absorption coefficients in the far-wing region of the $\nu_1+\nu_3$ (left panel) and $3\nu_3$ (right panel) bands. Results obtained using the ECS and Lorentzian models are compared with experimental data from  \citet{Burch_JOSA_1969, baranov1981_CO2He_edges, thibault1994_CO2He}. Results obtained by using Lorentzian profiles with a line-wing cutoff at 25~cm$^{-1}$ are also shown.}
  \label{fig.4}
\end{figure}

The ECS model and the corresponding set of parameters were subsequently used to calculate absorption in two additional spectral regions for which room-temperature experimental data are available, namely 3760--3940 and 6990--7050 cm$^{-1}$, corresponding to the far wings of the $\nu_1+\nu_3$ and $3\nu_3$ bands, respectively. For these calculations, contributions from all relevant bands were included using spectroscopic parameters from HITRAN \citep{gordon_hitran2020_2021} and pressure-broadening coefficients from \citet{HENDAOUI_CMDS_CO2He_2026}. As shown in Fig.~\ref{fig.4}, the ECS calculations are in very good agreement with the available experimental data in both spectral regions. These results further demonstrate the ability of the present ECS approach to accurately predict far-wing absorption of He-broadened CO$_2$ across different vibrational bands.

\section{$\chi$-factor line shape for He-broadened CO$_2$ lines}
\label{sec:chi}

For practical applications, such as opacity calculations for exoplanet atmosphere studies, we provide a set of empirical temperature- and wavenumber-dependent $\chi$-factor corrections to Lorentzian line shapes in order to account for the far-wing absorption of CO$_2$ broadened by He. The parameters of these corrections were fitted to far-wing absorption coefficients computed using the ECS model described above. Specifically, absorption spectra were calculated for four strong CO$_2$ bands, from the band centers to the far wings: the $\nu_2$, $\nu_3$, $\nu_1+\nu_3$, and $3\nu_3$ bands, centered near 15, 4.3, 2.7, and 1.4 $\mu$m, respectively. To reduce computational cost, only the contribution of the main band was accounted for in each spectral region. Calculations were performed for temperatures ranging from 200 K to 1000 K with a step of 100 K.

Within the $\chi$-correction model, the absorption of CO$_2$ infinitely diluted in He is expressed as:

\begin{align}
\alpha(\sigma, P, T) &= N_{\mathrm{CO_2}} 
\sum_l 
\frac{h c (\sigma - \sigma_l)}{k_B T} 
\frac{1}{1 - \exp\Big(-\frac{h c (\sigma - \sigma_l)}{k_B T}\Big)}
\nonumber\\[6pt]
&\quad \times
\frac{1 - \exp\Big(-\frac{h c \sigma}{k_B T}\Big)}
{1 - \exp\Big(-\frac{h c \sigma_l}{k_B T}\Big)}
\frac{\sigma}{\sigma_l}
\frac{1}{\pi}
\nonumber\\[6pt]
&\quad \times
\frac{S_l(T)\,\Gamma_l(P,T)}
{(\sigma-\sigma_l-\Delta_l(P,T))^2+(\Gamma_l(P,T))^2}
\chi(T, |\sigma-\sigma_l|)
\label{eq.8}
\end{align}
where $P$ is the He partial pressure, $N_{CO_2}$ is the CO$_2$ number density, and the sum extends over all absorption lines $l$ of position $\sigma_l$ contributing to the absorption at wavenumber $\sigma$. Here, $k_B$ is the Boltzmann constant, $S_l$ is the integrated line intensity, and $\Gamma_l$ and $\Delta_l$ are the pressure-induced half-width and line shift, respectively. We neglected here the contribution of CO$_2$-CO$_2$ collisions. For mixtures in which both CO$_2$--CO$_2$ and CO$_2$--He collisions contribute significantly, a more general formulation must be used (see \citealt{chaverot2025spect}). 

The factor $\frac{h c (\sigma - \sigma_l)}{k_B T} 
\frac{1}{1 - \exp\Big(-\frac{h c (\sigma - \sigma_l)}{k_B T}\Big)}$ is the quantum asymmetry factor, which accounts for the asymmetry between the low- and high-frequency far wings \citep{frommhold_collision-induced_1994, fakhardji_direct_2022, chaverot2025spect}. Finally, the $\chi(T, |\sigma-\sigma_l|)$ factors account for the sub-Lorentzian behavior in the line wings. Their analytical expression, similar to those used previously for self-, air-, and H$_2$O-broadened CO$_2$ \citep{perrin_temperature-dependent_1989, tran_measurements_2011,tran_co2broadened_2019, chaverot2025spect}, is given in table~\ref{Table2}.

\begin{table*}[t]
	\centering
    \caption{Analytical expression of $\chi$ factors}
     \label{Table2}
  
	\begin{tabular}{l p{9cm}}
        \hline  \hline 
        $\chi(T,\Delta\sigma)$  &  $\Delta\sigma$\\
         
        \hline 
        1 & $0\leq|\Delta\sigma|<\sigma_1$ with $\sigma_1=3$ cm$^{-1}$   \\
        $\exp[-B_1(|\Delta\sigma|-\sigma_1)]$ & $\sigma_1\leq|\Delta\sigma|<\sigma_2$ with $\sigma_2=30$ cm$^{-1}$ \\
        $\exp[-B_1(\sigma_2-\sigma_1)-B_2(|\Delta\sigma|-\sigma_2)]$ & $\sigma_2\leq|\Delta\sigma|<\sigma_3$ with $\sigma_3=180$ cm$^{-1}$ for the $\nu_2$, $\nu_3$ and $3\nu_3$ bands, while $\sigma_3=150$ cm$^{-1}$ for the $\nu_1+\nu_3$ band \\
          $\exp[-B_1(\sigma_2-\sigma_1)-B_2(\sigma_3-\sigma_2)-B_3(|\Delta\sigma|-\sigma_3)]$ & $\sigma_3\leq|\Delta\sigma|$\\
        
        \hline 
	\end{tabular}
 
\end{table*}

The parameters $B_i$ (table ~\ref{Table2}) are modeled as a function of temperature as: 

\begin{equation}
 B_i(T)=\alpha_i+\beta_i \exp\left(-\gamma_i \frac{T_0}{T}\right) 
 \label{eq.9}
\end{equation}

The parameters $\alpha_i$, $\beta_i$, and $\gamma_i$ were obtained by fitting Eqs.~\ref{eq.8} and \ref{eq.9} to the high-frequency wing absorption computed with the ECS model for each considered region and for nine temperatures between 200 K and 1000 K. The resulting values are listed in Table~\ref{table 3}.

\begin{table*}[t]
\centering
\caption{Parameters $\alpha$, $\beta$, and $\gamma$ for the four considered bands.}
\label{table 3}

\begin{tabular}{c c c c c c c c c c}
\hline \hline

& \multicolumn{3}{c}{$\nu_2$ band}
& \multicolumn{3}{c}{$\nu_3$ and $3\nu_3$ bands}
& \multicolumn{3}{c}{$\nu_1+\nu_3$ band} \\

\cline{2-4} \cline{5-7} \cline{8-10}

& $\alpha$ & $\beta$ & $\gamma$
& $\alpha$ & $\beta$ & $\gamma$
& $\alpha$ & $\beta$ & $\gamma$ \\

\hline

$B_1$ & 0.1832 & -0.0904 & 1.48
   & 0.1678 & -0.0850 & 2.94
   & 0.1520 & 0 & 0 \\

$B_2$ & 0.0235 & -0.0114 & 1.81
   & 0.0309 & -0.0167 & 0.47
   & 0.0279 & -0.0369 & 2.74 \\

$B_3$ & 0.0055 & 0 & 0
   & 0.009 & -0.0037 & 0.81
   & 0.0052 & 0.0006 & 1.10 \\

\hline \hline
\end{tabular}

\end{table*}

\begin{figure*}[h]
  \centering
  \includegraphics[width=\textwidth]{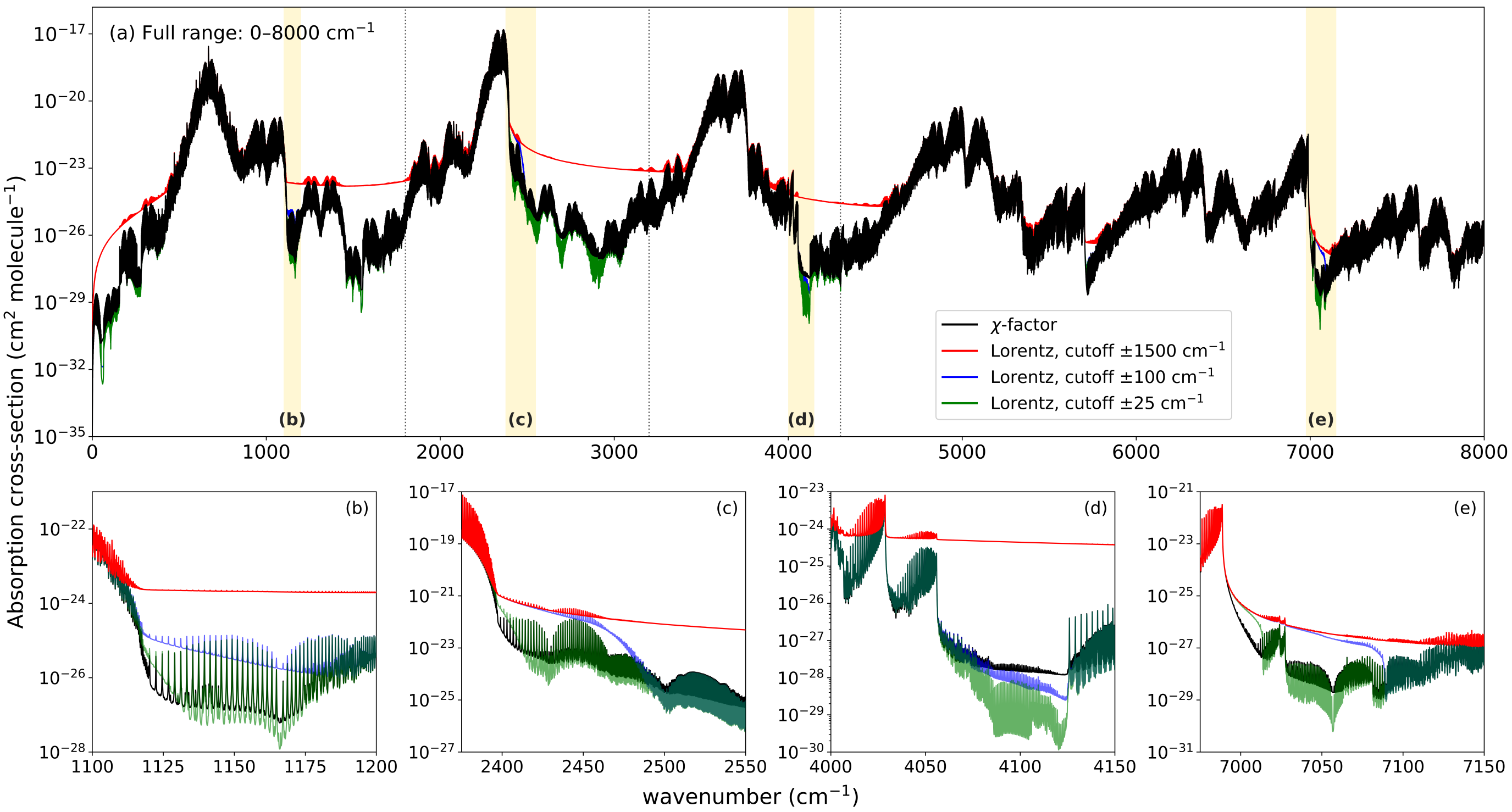}
  \caption{Comparison of He-broadened CO$_2$ absorption cross-sections at $P_{He}=1$ atm, $T=500$ K, calculated with and without (i.e. Lorentz, cutoff = $\pm~1500~cm^{-1}$) the $\chi$-factor corrections. Calculations using line-wing cutoffs of 25 and 100~cm$^{-1}$ are also shown. These values correspond to the cutoffs recommended for MAESTRO by \citet{gharib-nezhad_impact_2024} for pressures at or below 200~bar and above 200~bar, respectively. The He-broadening coefficients were taken from \citet{HENDAOUI_CMDS_CO2He_2026}, while the other spectroscopic parameters were taken from HITRAN \citep{gordon_hitran2020_2021}. The top panel presents the full spectral range from 0 to 8000 $cm^{-1}$, with vertical dotted lines indicating the spectral intervals associated with the four main CO$_2$ bands. The bottom panels show zoomed views of selected spectral regions.}
  \label{fig.5}
\end{figure*}

Figure~\ref{fig.5} compares infrared absorption cross-sections of CO$_2$ broadened by He computed with and without the $\chi$-factor corrections, at 500 K and 1 atm. Following the convention adopted in \citet{chaverot2025spect}, the $\chi$ factors derived from the $\nu_2$ band were applied to all lines located between 0 and 1800 cm$^{-1}$. Those derived from the $\nu_3$ band were applied between 1800 and 3200 cm$^{-1}$, while the $\nu_1+\nu_3$ parameters were used between 3200 and 4300 cm$^{-1}$. Above 4300 cm$^{-1}$, the $\chi$ factors derived from the $\nu_3$ band were again adopted. Note that applying a $\chi$-factor correction derived for a strong band to weaker bands within the same spectral region constitutes an approximation. To assess the uncertainty associated with this approximation, we performed a sensitivity analysis by varying the $\chi$ factor assigned to the weaker bands. First, we used the $\chi$ factor derived for the $\nu_3$ band for all weak bands, except for the $\nu_2$, $\nu_1+\nu_3$ and $3\nu_3$ bands, for which specific $\chi$ factors were determined. We then repeated the calculation by using, in turn, the $\chi$ factors derived for the $\nu_2$ and $\nu_1+\nu_3$ bands for all weak bands. The resulting spectra are very similar, with maximum relative differences remaining below 20~\% over the entire spectral range considered (0--8000 cm$^{-1}$). This indicates that the calculated spectrum is only weakly sensitive to the choice of $\chi$ factor for the weaker bands. We therefore retain the band-specific $\chi$-factor corrections reported in Table~\ref{table 3} as our best estimates for the corresponding spectral regions.  

As can be seen in Fig.~\ref{fig.5}, neglecting the $\chi$-factor corrections by using a purely Lorentzian line shape results in discrepancies exceeding three orders of magnitude in several spectral windows. When applying line-wing cutoffs of 25 cm$^{-1}$ or 100 cm$^{-1}$, as commonly done in opacity models for exoplanet atmospheres \citep{barstow2020comparison,chubb_exomolop_2021, gharib-nezhad_impact_2024}, the differences are reduced because the far-wing absorption is dominated by local lines. Nevertheless, differences of more than one order of magnitude remain in several spectral regions. These results clearly demonstrate the importance of realistic far-wing line-shape corrections for modeling CO$_2$ opacity in He-rich atmospheres. 

\section{Conclusions}
\label{sec:conclusions}
Far-wing absorption of CO$_2$ broadened by He is predicted for the first time using first-principles calculations. These calculations, based on requantized classical molecular dynamics simulations (rCMDS), involving no empirical parameters, provide the spectral density, or equivalently the band-normalized absorption spectrum, from the band centers to the far wings. All collisional effects including the two main mechanisms responsible for the sub-Lorentzian behavior observed in the far wings, namely line-mixing and non-impact effects, were accounted for in the rCMDS. Comparisons between the predicted far-wing absorption and experimental data in the far-wing region of the CO$_2$ $\nu_3$ band showed excellent agreement. Combined with previous studies demonstrating accurate predictions of pressure-broadening coefficients over a wide temperature range, these results confirm the ability of rCMDS to accurately describe the full range of collision effects affecting molecular absorption spectra.

Because non-impact effects are expected to be negligible for He-broadened CO$_2$ spectra and because rCMDS remain computationally demanding, available experimental data together with rCMDS predictions in the far-wing region of the $\nu_3$ band, for temperatures between 200 K and 1000 K, were subsequently used to constrain an impact-based spectral model that accounts for collisional line-mixing effects. The resulting model and associated parameter set provide an accurate representation of far-wing absorption in several spectral regions and can therefore be used to predict the far-wing absorption of He-broadened CO$_2$ throughout the infrared spectrum.

Using the far-wing absorption calculated with this model, we derived a set of temperature-dependent $\chi$-factor corrections that reproduce the sub-Lorentzian behavior of all major infrared CO$_2$ bands. These corrections provide a physically based set of parameters for modeling of the far-wing absorption of CO$_2$ broadened by He over a wide temperature range, in contrast to the commonly used line-wing cutoff approximations. They can be readily implemented in opacity calculations for CO$_2$ in He-rich atmospheres, such as those encountered in exoplanetary environments.

\begin{acknowledgements}
This work was granted access to the HPC resources of IDRIS under the allocation 2025-[A0180914906] made by GENCI. It was also supported by the Programme National de Planétologie (PNP) of CNRS-INSU co-funded by CNES.
\end{acknowledgements}

\bibliographystyle{aa}
\bibliography{biblio}
   
\end{document}